\documentclass{aa}
\usepackage{graphicx,amsmath,fixltx2e}
\usepackage{txfonts,color}
\usepackage{hyperref}
\usepackage{natbib}
\usepackage{textcomp}
\usepackage{placeins}
\bibpunct{(}{)}{;}{a}{}{,}

\begin{document}
\title{Properties of sunspot umbrae observed in Cycle 24}

\author{Christoph Kiess
  \and Reza Rezaei
  \and Wolfgang Schmidt}
  \authorrunning{Kiess, Rezaei, \& Schmidt}
\institute{Kiepenheuer-Institut f\"ur Sonnenphysik, Sch\"oneckstr. 6, D-7104 Freiburg, Germany\\
\email{[kiess, rrezaei, wolfgang]@kis.uni-freiburg.de}}

\abstract
{}
{There is an ongoing debate whether the solar activity cycle is overlaid with a long-term decline that may lead to another grand minimum in the near future. 
We use the size, intensity, and magnetic field strength of sunspot umbrae to compare the present cycle 24 with the previous one.}
{
We used data of the Helioseismic and Magnetic Imager onboard the Solar Dynamics Observatory and selected all
sunspots between May 2010 and October 2012, using one image per day. 
We created two subsets of this data with a manual tracking algorithm, both without duplication. 
One is containing each sunspot (910 umbrae within 488 spots) and was used to analyze the distribution of umbral areas, selected with an automated thresholding method. 
The other one contains 205 fully evolved sunspots. We estimated the magnetic field and the total magnetic flux 
or those and discuss the relations between umbral size, minimum continuum intensity, maximum field strength and total magnetic flux.
}
{
We find nonlinear relations between umbral minimum intensity and size and between maximum magnetic field strength and size. The field strength scales linear with the intensity and the umbral size scales roughly linear with the total magnetic flux, while the size and field strength level off with stronger flux. 
When separated in hemisphere and averaged temporally, the southern umbrae show a temporal increase in size and the northern umbrae stay constant. 
There is no temporal variation in the umbral mean intensity detectable.
The probability density function of the umbral area in the ascending phase of the current solar cycle is similar to that of the last solar cycle.
}
{
From our investigation of umbral area, magnetic field, magnetic flux and umbral intensity of the sunspots of the rising phase of cycle 24, 
we do not find a significant difference to the previous cycle, and hence no indication for a long-term decline of solar activity.
}

\maketitle

\section{Introduction}
Sunspots are the most conspicuous manifestations of the solar magnetic activity in white light images,  
\citep{bray_loughhead_64, wittamann_xu_1987}. They are at the top of the hierarchy of the photospheric 
magnetic structures, being about three orders of magnitude larger than magnetic flux tubes 
\citep{keller_1992, wiehr_etal_2004, rouppe_etal_2005, solanki_etal_06,lagg_etal_2010}. 
Diameters of sunspot umbrae span a range between 1 to 20 Mm and their areas cover two orders of magnitude 
\citep{solanki03r,rempel_schlichenmaier_2011,borrero_ichimoto_2011}. 
The umbrae of the largest sunspots are the coolest structures 
observed in the solar photosphere, their temperature lies some 2000\,K below the quiet Sun photosphere \citep{maltby_etal_86, kopp_rabin_92, balthasar_schmidt_1993, martinez_vazquez_1993, collados_etal_94}. 
The orientation of the magnetic field is close to the local vertical in umbrae \citep{beckers_schroeter_1969,keppen_martinez_96,mathew_etal_2003, beck2008}. 

The periodic variation of sunspot number, i.e., the solar cycle, 
was observed over 400 years \citep{athay_warwick_1961, harvey_1992, spruit_2012, usoskin_2013}. 
It is traced with several quantities including the number and area of 
sunspots \citep{hoyt_schatten_1996, hathaway_etal_2002, usoskin_2008, hathaway_2010}. 
The cyclic variation of umbral properties like intensity and magnetic field strength was subject of several investigations 
\citep{livingston_02, penn_livi_06, penn_macdonald_07, penn_livi_10,  livingston_etal_2012, pevtsov_etal_2013}. 
Among others, \cite{albregtsen78} and \cite{albergtson_etal_1984} find that the umbral intensity fluctuates in phase
with the solar cycle such that the umbrae are brighter at the end of a cycle. \cite{norton_gilman_2004} find a decrease in the umbral intensity 
from early to mid phase of cycle 23 and an increase after the maximum of the cycle. In contrast, \cite{mathew_etal_2007} report an invariant umbral intensity 
through the cycle 23.
This is accompanied with a periodic variation of the maximum field strength of umbrae with the solar 
cycle phase \citep{watson_etal_2011, pevtsov_etal_2011, reza_etal_2012a}.

The fraction of the umbral area to the sunspot area is apparently independent of the solar cycle. Using Greenwich photoheliographic results, 
\cite{jensen_etal_1955}, \cite{tandberg_1956} and \cite{ringnes_1964} report a weak tendency 
for the average penumbra to umbra radius to change in phase with the solar cycle. 
\cite{steinegger_etal_1990} and \cite{brandt_etal_1990} find umbra-penumbra area ratios of 0.24 and 0.32 for small and large spots, respectively.
\cite{beck_chapman_1993} find the umbra-penumbra area ratio independent of the sunspot complexity 
and magnetic field strength and only weakly correlated with the umbral area. 
\cite{mathew_etal_2007} find no variation of the umbral radius with the solar cycle. 
They note, however, a weak tendency of a secular trend in the northern hemisphere. 
\cite{penn_macdonald_07} and \cite{reza_etal_2012a} find no significant variation of the umbral area as a function of the solar cycle. 
\citet[][hereafter SP]{schad_penn_10} observe a small variation in the umbral size correlated with 
the variation of the umbral intensity and the solar cycle.

The general view in old observations was that smaller sunspots have a higher continuum intensity and 
a lower magnetic field strength \citep{bray_loughhead_64}. 
The underestimation of stray light in earlier works was noticed by \cite{zwaan_1965}, who argued that much of the dependency of the 
continuum intensity on the sunspot size can be explained by stray light \citep[see also][]{rossbach_schroeter_1970}. More recent investigations 
challenged the idea that the continuum-area dependency is an artifact of stray light \citep{mcintosh_1981,stellmacher_wiehr_1988,martinez_vazquez_1993}. 
Most recent observations indicate that the continuum-area dependency is a 
real one and exists even after removal of stray light \citep{mathew_etal_2007, wesolowski_etal_08, schad_penn_10, reza_etal_2012a}.

Measurements of umbral physical properties are prone to systematic and random errors. 
Intensity measurements of umbrae are affected by scattered light \citep{mattig1971, martinez_pilet_1992, chae_etal_1998, beck_reza_2011a}. 
The field strength measurements in sunspot umbrae using the Zeeman splitting of spectral lines 
are contaminated on the one hand by molecular blends \citep{wittmann_1972} and on the other hand by 
inaccurate atomic data \citep{borrero_etal_2003}. 
The uncertainty in the measurements of the umbral area is due to the umbral fine 
structures \citep{sobotka_etal93,schmidt_balthasar_94,lites_etal_04, rimmele_08, shimizu_etal_09}. 
Umbrae are more stable compared to penumbrae \citep{robinson_boice_82,leka_skumanich_1998,rolf_etal_2010a, reza_etal_2012b, louis_etal_2012}.
Larger spots tend to have darker umbrae, higher magnetic field strength, and longer lifetime compared to smaller spots \citep[e.g.,][]{schrijver_1987}. 
However, there is a significant intrinsic scatter \citep{lites_skumanich_1990,martinez_vazquez_1993,schlichenmaier_collados_2002}.

\citet[][hereafter BOG]{bogdan_etal_1988} and SP study the umbral size distribution 
from 1917 to 1982, and during the solar cycle 23, respectively. 
They find that there is no significant temporal variation in the size distribution from one cycle to the other.  
Such an invariance in the umbral area distribution is noteworthy considering the drastic difference in the number of sunspots in  
different solar cycles. The distribution can be fitted with a lognormal function, 
motivated by the fragmentation and random processes relevant to 
sunspot formation \citep[][and references therein]{bogdan_etal_1988}. 
The size distribution of sunspot groups in the Greenwich photoheliographic sunspot records was studied by 
\cite{baumann_sol_2005} who find a lognormal distribution both for 
the instantaneous area (corresponding to BOG selection) and maximum area of sunspots. 

The end product of a multiplicative and fragmentary process is a lognormal distribution \citep{kolmogorov}.
There are many applications of lognormal distributions in different branches of 
science \citep[e.g.,][]{pauluhn_etal_2000, cme_lognormal, kobayashi_etal_2011}. 
In the photosphere, a lognormal distribution of the umbral area observed in about three orders of magnitude 
hints at a fragmentation origin for umbrae. 
BOG speculate that 
a lognormal distribution of emerging flux tubes indicates that they are probably the result of fragmentation 
of a large flux tube. As pointed out by \cite{baumann_sol_2005}, it is unlikely to imagine that sunspots 
are purely the result of such fragmentation. It is perhaps more reasonable to imagine that they are  joint product of a fragmentation and 
coalescence of flux patches as discussed by \citet{zwaan_1992} and \citet{rolf_etal_2010a}. 

Such a size distribution provides insights in the 
nature of the solar dynamo mechanism \citep{parker_1979, ossendrij_2003}. 
Note however that most of the observed umbrae by BOG and SP were in their 
decay phase, so one should be careful about using their distribution as an argument for the origin of sunspot flux trunks in the convection zone. 
 The fragmentation of the rising flux tube also happens near the solar surface \citep{zwaan_1978, brandenburg_2005}. 
In a 16\,Mm deep domain, \cite{rempel_2011} finds fragmentation of the flux trunk down to the bottom boundary. 
This subsurface fragmentation causes fragmentation in the photosphere as observed by e.g., \cite{louis_etal_2012}.
To distinguish between two sunspot formation scenarios (fragmentation or coalescence), 
one has to evaluate the distribution in an early stage of spot development, when the decay has not redistributed the surfaced flux. 
It would be interesting to evaluate the umbral area beneath the photosphere and prior to their photospheric appearance. 
If lognormal, then it hints at a fragmentation process in the convection zone. 
To get as close as possible to pre-emergence situation, 
we have evaluated the umbral area distribution in an early stage, 
when the decay has not redistributed the surfaced flux significantly (Sectoral.\,\ref{sec:selection}).

Magnetohydrodynamic simulations of emerging flux tubes \citep{fan_2008}, 
active regions \citep{cheung_etal10}, pores \citep{cameron_etal_2007}, 
and sunspots \citep{rempel_etal_09} became recently available. 
The flux emergence simulations model the rise of a twisted flux tube in the convection zone, its expansion, and 
fragments near the solar surface. 
In contrast in sunspot and pore simulations, the field lines are imposed on a relaxed hydrodynamic domain. 
Then, the dispersed flux elements partially evacuate and merge \citep{kitiashvili_etal_2010} 
to form pores and sunspots. 
The coalescence of the emerged flux elements in the simulations is in agreement with 
observations of e.g., \cite{bernasconi_etal_2002} and \cite{rolf_etal_2010b}. 
The tethered-balloon model of \cite{spruit_1981} predicts that the coalescence of small flux patches \citep{pariat_etal_2004, reza_etal_2012a} is
a surface phenomenon since these flux elements are deeply anchored. In other words, the emergence of
large flux tubes does not fundamentally affect their integrity. 

Due to the large range of parameters, 
simulations usually focus on some aspects of sunspot properties rather than addressing all issues.
Umbrae in these simulations perform convective energy transport by umbral dots and transient light-bridges. 
\citet[][his Fig.\,11]{rempel_2011} presents a sunspot with a minimum (bolometric) umbral intensity of about 0.3\,I$_{\mathrm{c}}$. 
In another simulation he finds a minimum umbral intensity of about 0.22\,I$_{\mathrm{c}}$ \citep[][his Fig.\,5]{rempel_2012}. 
In both case, the unsigned flux of the simulated spot is about 10$^{22}$\,Mx. 
Using the SIR code \citep{sir92, sir_luis}, we performed spectral synthesis of the \citet{rempel_2012} umbra and find a 
a minimum continuum intensity of 0.08\,I$_{\mathrm{c}}$ at a wavelength of 630\,nm. 
This minimum intensity is close to what one expects for such a big sunspot (compare  with the sunspot of Nov 19, 2013).

\cite{rempel_2011} presents a simulated sunspot in a 16\,Mm deep domain. In this simulation, a considerable amount of 
subsurface magnetic field is fragmented due to convection. The photospheric manifestation of this fragmentation 
is the appearance of light-bridges or flux separation. 
He finds a deep-reaching outflow beneath the photosphere and down to the bottom boundary of the simulation 
domain.
Such an axisymmetric outflow presumably supports the monolithic subsurface structure of sunspots \citep{parker_1979a}. 
It is not clear what role the supergranulation flow plays in this process. 

Sunspots decay  as a result of fragmentation \citep{Petrovay_1997, valentin_2002}. 
The decay rate, which in turn determines the sunspot lifetime, 
is a function of anchoring depth and the corresponding convection timescales in such depths \citep{moradi_etal_2010}. 
The life time of sunspots, $T$, falls in-between two extremes:
$$\frac{H_{\mathrm{p}}}{\it{v}_{\mathrm{rms}}} \ll T \ll \left(\frac{r_0}{10\,\mathrm{Mm}}\right)^2 \times10\,[\mathrm{days}],$$
where $H_{\mathrm{p}}$ is the pressure scale height, $\it{v}_{\mathrm{rms}}$ is a typical convection velocity,
and $r_\mathrm{0}$ the maximal radius of the spot \citep{petrovay_moreno_1997, schlichenmaier_1999}.

\cite{mcintosh_1981} reports on a long-lived sunspot in Aug 1966, which survived five solar rotations (137 days).
While the convection timescale (the lower limit) is about 6\,hr at a depth of 
15\,Mm, \cite{rempel_2011} speculates that the lifetime of a sunspot is roughly an order of magnitude longer 
than the convection timescale (at the corresponding anchoring depth). 
The trend found in his simulation cannot be extrapolated to broader ranges since the aforementioned long-lived sunspot will need 
an anchoring depth of about four times the depth of the convection zone.

The prolonged duration of the minimum of the last solar cycle  (23) raises speculations about a lower activity level 
in solar cycle 24 and the possible approach of a new grand minimum \citep{tripathy+etal2010,jain+etal2011,tap_etal_2011}. 
There are also indications that the population of small sunspots compared to the large ones in  
solar cycle 23 might be different compared to previous cycles \citep{clette_lefevre_12}. 
This motivated us to study properties of umbrae in the current solar cycle (24) 
and compare it with previous cycles.

In this contribution, we measure the umbral area, evaluate the maximum field strength and 
minimum intensity of sunspot umbrae in solar cycle 24 based on the data obtained with 
the Helioseismic and Magnetic Imager \citep[HMI,][]{scherrer_etal_2012, hmi_2012}, a filter instrument 
onboard the Solar Dynamics Observatory \citep[SDO,][]{sdo}. 
From May 2010 till October 2012, 4229 sunspots were observed with some 6892 (910 unique)  umbrae. 
The criteria for the data selection are discussed in Sect.\,2. We analyze the data and interpret the results in 
Sects.\,3 and 4, respectively. A summary and conclusions are presented in Sect.\,5.

\section{Observations and data analysis}
HMI records six equidistant wavelength positions around the center wavelength of the 
neutral iron line at 617.33\,nm \citep{norton_etal_2006,fleck_etal_2011}. 
HMI filters have a full width at half maximum (FWHM) of about 7.6\,pm. 
With a spatial sampling of about $0.5\arcsec$, the spatial resolution of each full disk image is about $1\arcsec$. 
The image size is  $4096 \times 4096$ pixels and the cadence of full Stokes maps is 12\,min. \footnote{Data is available 
at \href{http://jsoc.stanford.edu/}{http://jsoc.stanford.edu/}, 
see also \href{http://hmi.stanford.edu/}{http://hmi.stanford.edu/} 
for information about the instrument.} 

\subsection{Range of observation}
\begin{figure}
\resizebox{\hsize}{!}{\includegraphics[width = \linewidth]{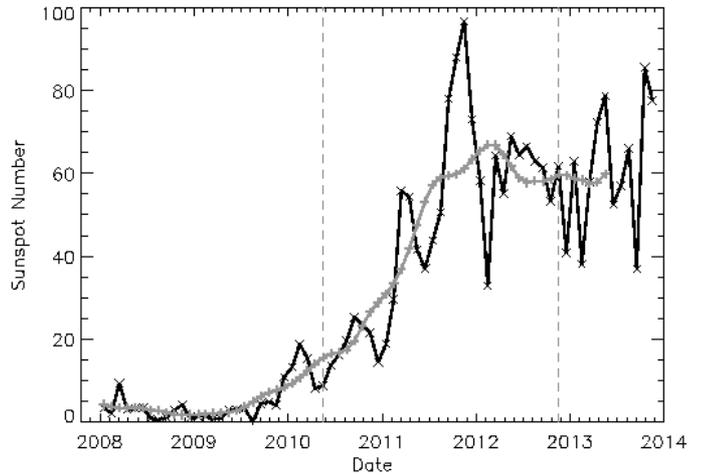}}
\caption{SIDC monthly sunspot number (black line). The gray line shows the smoothed sunspot number.
The dashed vertical lines mark the range of our data.
}
\label{fig:sidc}
\end{figure}

The current solar cycle (24) started on January 04, 2008, when a new active region with reversed polarity in the northern hemisphere at a high latitude 
was observed by the Solar and Heliospheric Observatory \citep[SOHO,][]{soho} Michelson Doppler Imager \citep[MDI,][]{scherrer_etal_1995}. 
Figure\,\ref{fig:sidc} shows the monthly (black line) and smoothed (gray line) sunspot number in cycle 24 \citep{sidc}.
The activity level was very low through the rest of 2008. A significant rise of the sunspot number only occurred  in early 2009 (Fig.\,\ref{fig:sidc}). 

Our observations (01/05/2010 to 31/10/2012) cover the rising phase of cycle 24. After a small dip in early 2013, the spot number increases again in the second quarter of 2013. 
We analyzed one image of the continuum intensity per day 
and selected all sunspots in the data.

\subsection{Correction for limb-darkening}
As a first step, each image was normalized to its mean value, derived from a $100\arcsec$ square at disk center, avoiding sunspots. 
To derive the limb-darkening curve, images from twelve days with no activity were chosen, and $\mu$, the cosine of 
the heliocentric angle, was computed for each pixel.
We used a smoothed version of the intensity as a function of $\mu$ to correct for limb-darkening.
This curve represents our empirical limb-darkening function. For each data pixel, the heliocentric angle was 
calculated and the intensity was divided by the corresponding value of our limb-darkening function. 

\subsection{Selection of umbrae} 
\label{sec:selection} 
We analyzed one image per day and manually marked a total of 4229 sunspots (6892 umbrae) on the visible solar hemisphere. 
Since most sunspots live for several days 
we applied an algorithm to reduce the data such that no sunspot is counted multiple times.

Starting with the first spot in our data, we 
calculated its expected positions during its passage across the solar disk (maximum 14 days), taking the differential rotation depending 
on the latitude $\theta$ into account. We used the empirical formula $\Omega~[\textrm{deg / day}] = 14.522 - 2.84 \sin ^2(\theta)$ given by
\citet{howard_1984}.
For those 14 days we then simultaneously plotted all sunspots closer than 15 degrees to the expected positions of the first spot. 
From this series of images 
we then manually selected the sunspots of interest and skipped the rest. 
This algorithm was successively used on the remaining sunspots in our original dataset (4229 spots).

We applied the described algorithm using two different selection criteria. For the first subset of data we focused on young spots with well developed umbrae. None of the selected spots had a heliocentric angle greater than 60 degrees ($\mu \ge 0.5$). 
This subset containing 488 spots was used for statistical analysis of umbral sizes described in Sect.\,\ref{sec:irb}. 
For the second subset we focused on fully evolved sunspots with penumbrae. In order to reduce scatter, no complex spots were selected. 
This led to a subset of 205 sunspots that was used to analyze the correlation of umbral properties.

\begin{figure}
\resizebox{\hsize}{!}{\includegraphics*[width = \linewidth]{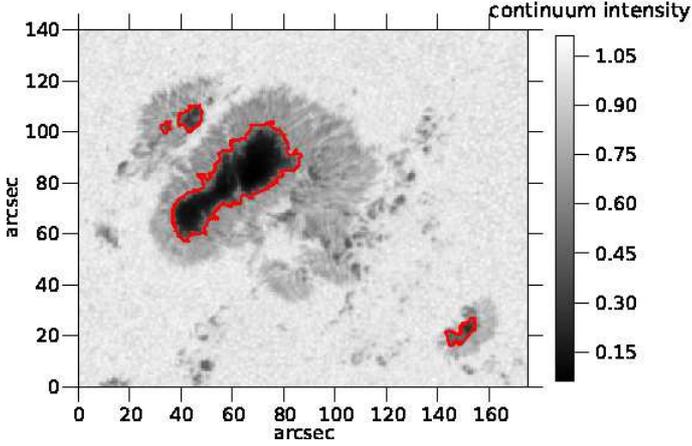}}
\caption{AR11520 on 14/07/2012 is an example of a sunspot with a large umbra. The red contour shows the threshold level 
(0.6 of mean quiet Sun intensity) used to select the umbra. 
The spot has a heliocentric angle of $\mu\,=\,0.86$. The total area of the largest four umbrae is 206\,MHS.} 
\label{fig:big_umbra}
\end{figure}

\subsubsection{The large sample}
The first subset (488 sunspots) was used for an analysis of the distribution of umbral sizes. We applied a threshold method to select the 
umbrae, see e.g., \cite{brandt_etal_1990},  \cite{mathew_etal_2007}, \cite{wesolowski_etal_08}, and SP.
This threshold was chosen as 0.6 of the mean quiet Sun intensity. The minimum size for an umbra to be selected was chosen to be 
$0.5$\,microhemispheres (MHS). $1~\textrm{MHS}$ is $10^{-6}$ of the visible solar surface and corresponds to $3.05~\textrm{Mm}^2$ or 
$5.79~\textrm{arcsec}^2$.
We counted distinct umbrae within one sunspots separately which led to a total of 910 umbrae within 488 sunspots. 
A threshold value of 0.6 ensures that all umbrae are sampled 
while it does not capture {\it umbral light bridges} as seen in Fig.\,\ref{fig:big_umbra} \citep{muller_1979}. 
To evaluate the umbral size distribution we created a histogram with fifty bins.
A logarithmic bin size was used to assure that there was still a fair number of umbrae in the bins covering larger values of the size spectrum.
We considered an uncertainty of $\pm1$ pixel for the umbral radius in each measurement. 
The relative uncertainty of the umbral size decreases with increasing umbral area. 
Since logarithmic binning was chosen, the bins for small spots are narrower than the bins for large ones. 
For the first bin ($0.5$\,MHS) the width is about the same as the uncertainty of the umbral area.

\subsubsection{The small sample}
For each of the 205 spots we used a threshold of 0.52 of the mean quiet Sun intensity to select its umbra 
(see Sect.\,\ref{sec:threshold} for the influence of the intensity threshold).
From a circle with the same area 
as the umbra we calculated an equivalent radius. We also derived the minimum relative intensity (the darkest pixel) and the maximum magnetic field strength. We further marked the penumbra manually and calculated the total magnetic flux of the sunspot.
Using this sample, 
we studied relations between size, intensity, field strength and flux of sunspots. 
We also analyzed the variation of minimum intensity and size measurements in each hemisphere as a function of time. 
We group this data using ten measurements in each but the last bin.
The radii in the large and the small sample are not directly comparable, since we applied 
different thresholds, and in the small sample we did not distinguish between multiple umbrae within one sunspot.

\subsection{Inversion of HMI data}\label{sec:inversion}
\begin{figure}
\resizebox{\hsize}{!}{\includegraphics*[width = \linewidth]{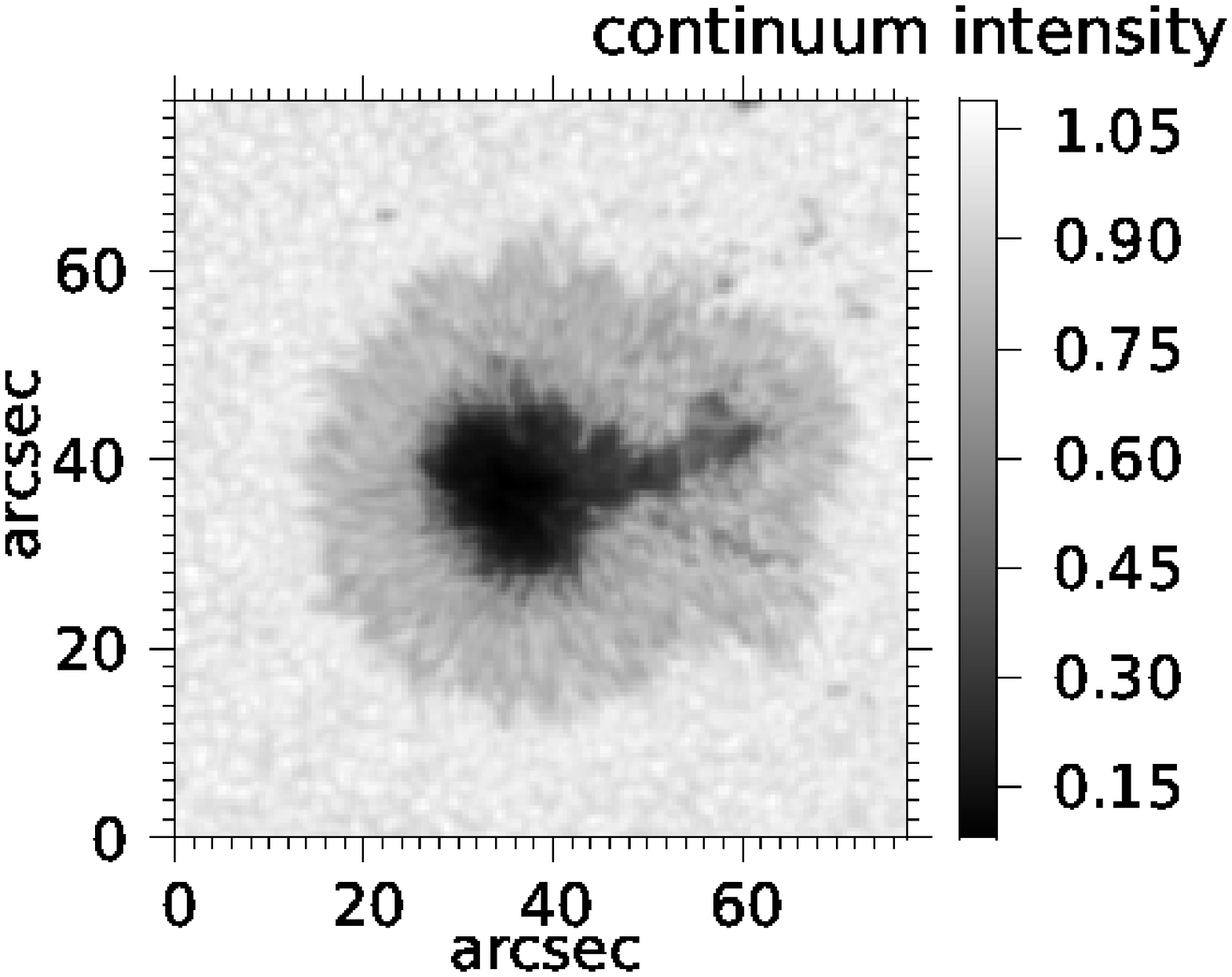}\includegraphics*[width = \linewidth]{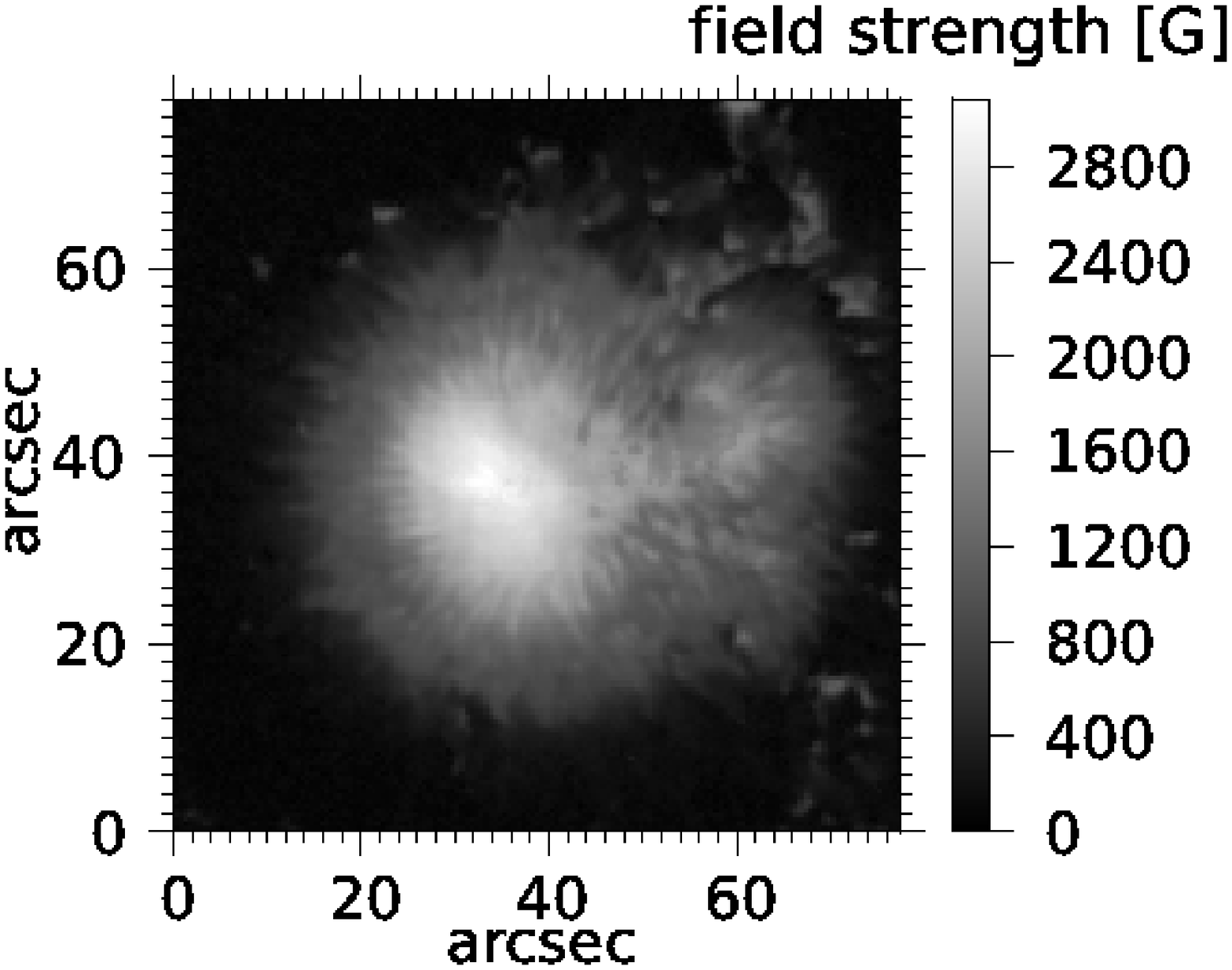}}
\resizebox{\hsize}{!}{\includegraphics*[width = \linewidth]{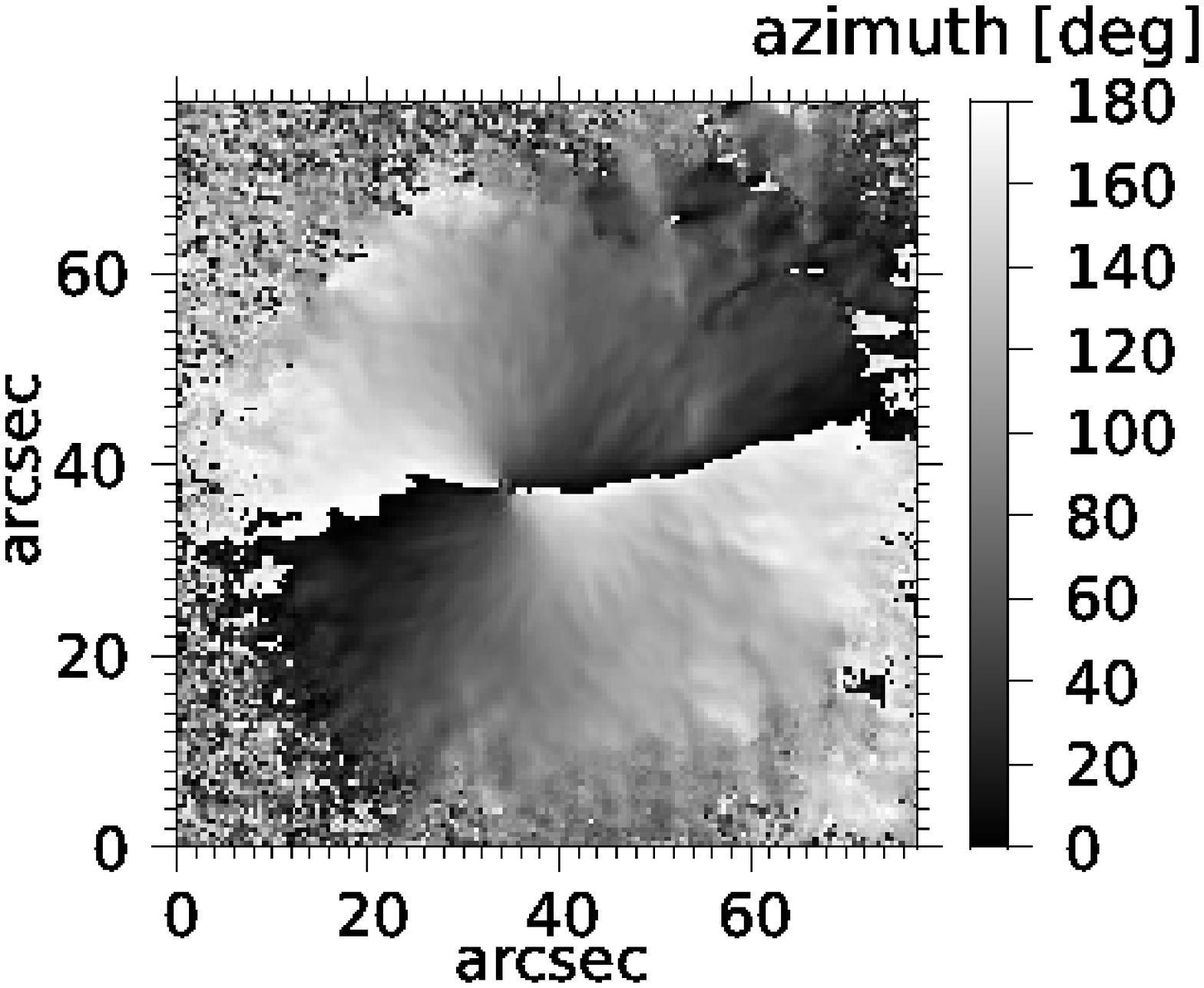}\includegraphics*[width = \linewidth]{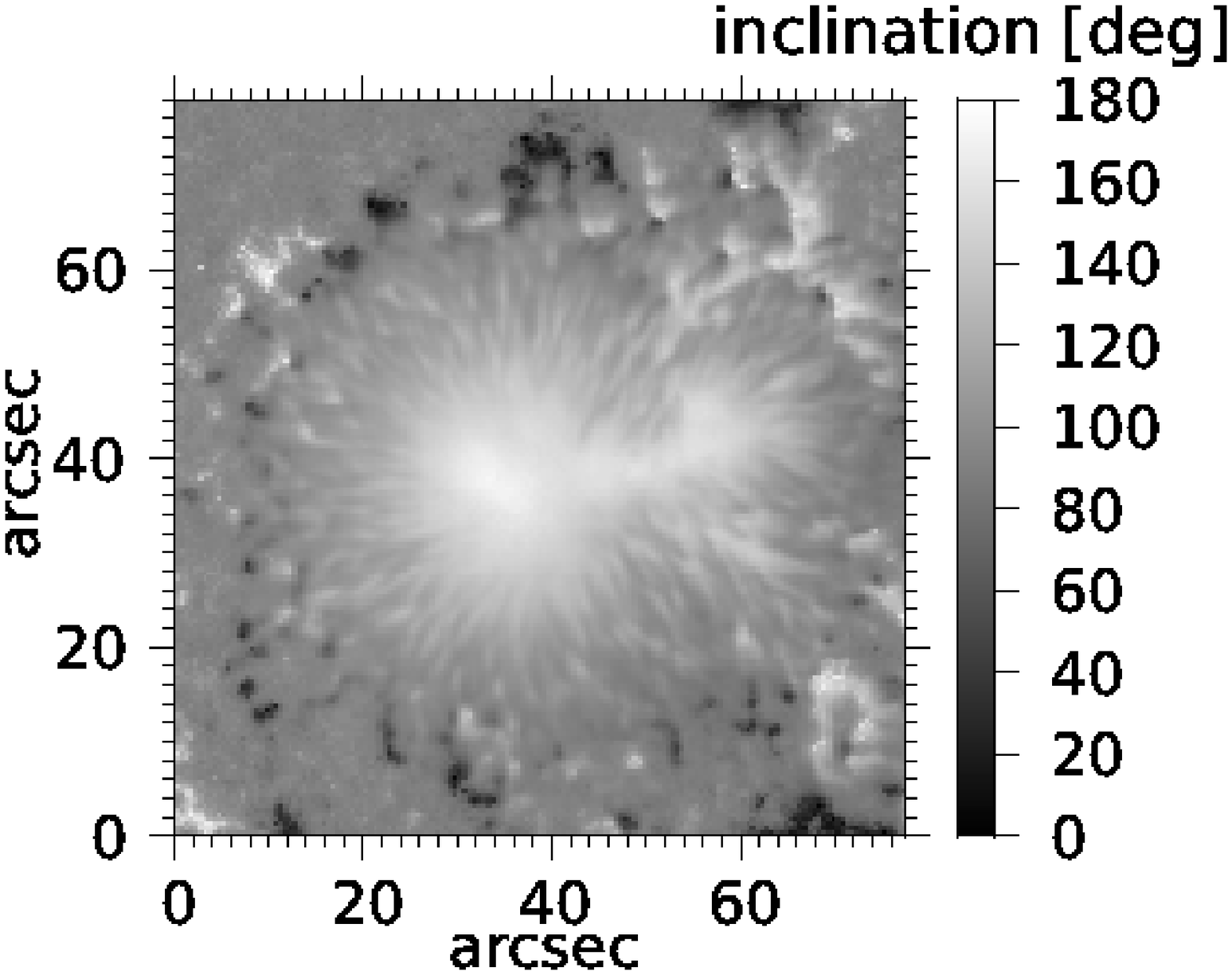}}
\caption{
Inversion results for one of our sunspots. From top left clockwise: continuum intensity, 
magnetic field strength, inclination, and azimuth.
}
\label{fig:inversion}
\end{figure}

We performed an inversion of HMI data using an updated version of the Very Fast Inversion of the Stokes Vector \citep[VFISV,][]{borrero_vfisv_2007, borrero_vfisv_2011} code 
to retrieve the magnetic field. 
Assuming a source function that varies linearly with optical depth, the code uses a 
Milne-Eddington approximation to solve the radiative transfer equation. The best-fit solution of the code then returns the 
height-independent values of the magnetic field strength, inclination and azimuth along with the velocity and other parameters. 
The inversion was performed for all sunspots of the small sample. An example of the results 
is shown in Fig.\,\ref{fig:inversion}. 
The standard deviation of the estimated error of the magnetic field strength is about 87 Gauss (G). 
Since HMI samples only six wavelength positions systematic errors in the magnetic field strength are possible.
Although the magnetic field strength in magnetograms suffers from saturation effects \citep{liu_norton_scherrer_2007, liu_hoeksema_etal_2012}, 
we do not see saturation in the inverted field strength. There are, however, limitations on the maximum retrieved field strength using HMI data. 
This is mainly due to low light level in the center of dark umbrae, the spacecraft velocity, presence of molecular blends, as well as large Zeeman 
splitting  compared to the spectral range of HMI. 
We did not use the spurious field strength, which appeared only on a few large sunspots in our sample, but used the maximum field strength where the inversion succeeded.


\section{Results}

\begin{figure}
\resizebox{\hsize}{!}{\includegraphics[width = \linewidth]{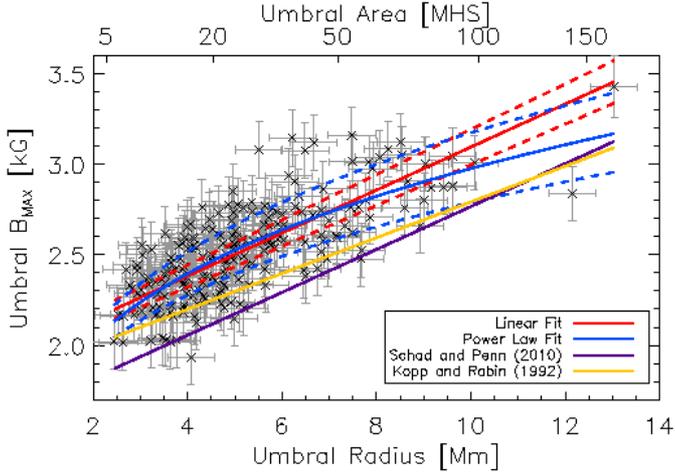}}
\caption{
Variation of umbral maximum field strength as a function of radius. Parameters of a linear fit (solid-red line) and power-law fit (solid-blue line) and correlation coefficients are given in Table\,\ref{tab:result1}. 
The dashed lines mark the one sigma confidence level for each fit. 
The yellow and purple lines mark the adopted linear fits from \cite{kopp_rabin_92} and SP, respectively.
The upper $x$-axis shows the area corresponding to the given radius.
}
\label{fig:bmax_size}
\end{figure}

\begin{figure}
\resizebox{\hsize}{!}{\includegraphics[width = \linewidth]{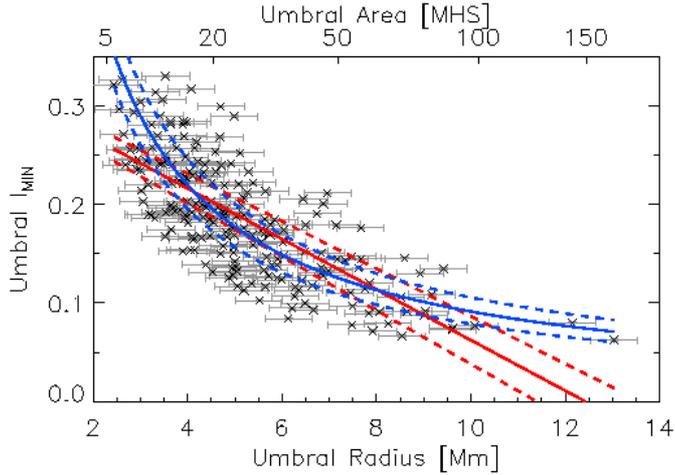}}
\caption{
Variation of umbral minimum intensity vs. radius. 
For more description, see caption of Fig.\,\ref{fig:bmax_size}.
}
\label{fig:imin_size}
\end{figure}

\begin{figure}
\resizebox{\hsize}{!}{\includegraphics[width = \linewidth]{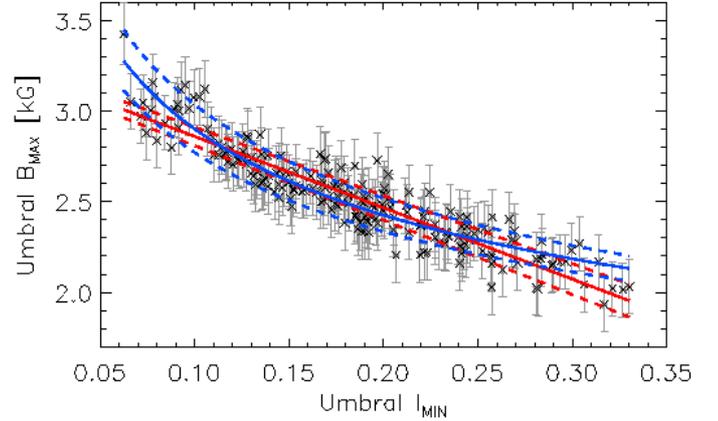}}
\caption{
Maximum field strength of umbra vs. minimum intensity.
For more description, see caption of Fig.\,\ref{fig:bmax_size}.
}
\label{fig:imin_b}
\end{figure}

\subsection{Size, intensity, \& field strength}\label{sec:irb}
We study the umbral maximum field strength, minimum continuum intensity, and area (or its equivalent radius) in our data. 
To this end, we selected 205 mature sunspots not far from disk center, with relatively simple structure, and roughly after the end of their
initial growth phase. Figure\,\ref{fig:inversion} shows a typical spot of this sample. 

Figures\,\ref{fig:bmax_size} and \ref{fig:imin_size} 
show scatter plots of  maximum field strength and minimum intensity as a function of umbral radius, respectively. 
Figure\,\ref{fig:imin_b} shows the relation between field strength and intensity. We plot power-law (blue) and linear fits in all three cases. Fit parameters can be found in Table\,\ref{tab:result1}. 

A power-law fit (blue lines) reproduces the observed data better than a linear fit in Figs.\,\ref{fig:bmax_size} and \ref{fig:imin_size}. 
This nonlinearity motivated us to fit a power-law in the intensity-magnetic field strength relation (Fig.\,\ref{fig:imin_b}) as well.
The three power exponents are consistent, i.e., 0.23\,$\approx$\,(-0.96)\,*\,(-0.26), cf. Table\,\ref{tab:result1}.
We also provide linear fits for comparison with other authors (see Sect.\,\ref{sec:discussion}).
\begin{table*}
\centering
\caption{
Fit parameters and one sigma errors for the linear ($y = \textrm{A} + \textrm{B}x$) and power-law ($y = \textrm{C}  * x^\textrm{D}$) relations for Figs.\,\ref{fig:bmax_size}\,\--\,\ref{fig:flux}. $R$ is the umbral equivalent radius in Mm, $I$ the relative minimum intensity, $B$ the maximum field strength in kG and and $\Phi$ the total magnetic flux in $10^{21}$\,Mx. C$_\textrm{P}$ and C$_\textrm{S}$ are the Pearson and Spearmann's correlation coefficients, respectively.
}
\begin{tabular}[h]{cccccccrr}
\hline \hline
\multicolumn{1}{c}{Fig.} & \multicolumn{1}{c}{x} & \multicolumn{1}{c}{y} & \multicolumn{1}{c}{A} & 
\multicolumn{1}{c}{B}  & \multicolumn{1}{c}{C} & \multicolumn{1}{c}{D} & \multicolumn{1}{c}{C$_\textrm{P}$} 
& \multicolumn{1}{c}{C$_\textrm{S}$} \\
\hline
\ref{fig:bmax_size} & $R$ & $B$ & $1908\pm35$ &  $119\pm6$ & $1726\pm48$ & $0.24\pm0.02$ & $0.73$ & $0.74$ \\
\ref{fig:imin_size} & $R$ & $I$ & $0.319\pm0.009$ & $ -(2.56\pm0.16)10^{-2}$ & $0.830\pm0.054$ & $-0.958\pm0.035$  & $-0.78$ & $-0.76$ \\
\ref{fig:imin_b} & $I$ & $B$ & $3254\pm34$ & $-3937\pm171$ & $1601\pm32$ & $-0.26\pm0.01$  & $-0.92$ & $-0.93$ \\
\ref{fig:flux} & $\Phi$ & $B$ & - & - & $2176\pm30$ & $0.126\pm0.009$  & $0.67$ & $0.67$ \\
\ref{fig:flux} & $\Phi$ & $R$ & $2.36\pm0.07$ & $0.763\pm0.019$ & - & -  & $0.96$ & $0.95$ \\
\ref{fig:flux} & $\Phi$ & $I$ & - & - & $0.368\pm0.012$ & $-0.586\pm0.017$  & $-0.72$ & $-0.68$ \\
\hline
\end{tabular}
\label{tab:result1}
\end{table*}

\begin{figure*}[h!]
  \resizebox{\hsize}{!}{\includegraphics[width = \linewidth]{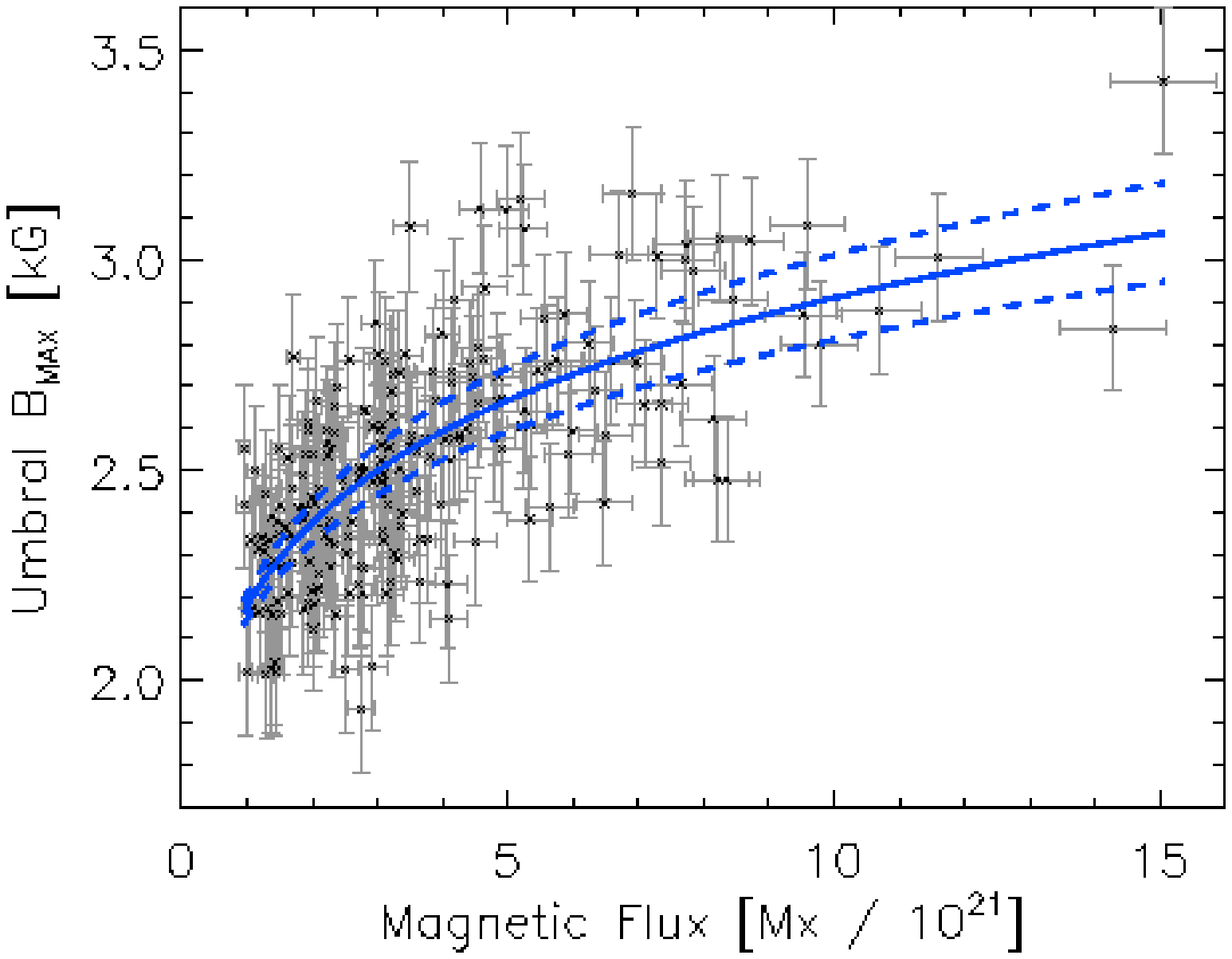}
  \includegraphics[width = \linewidth]{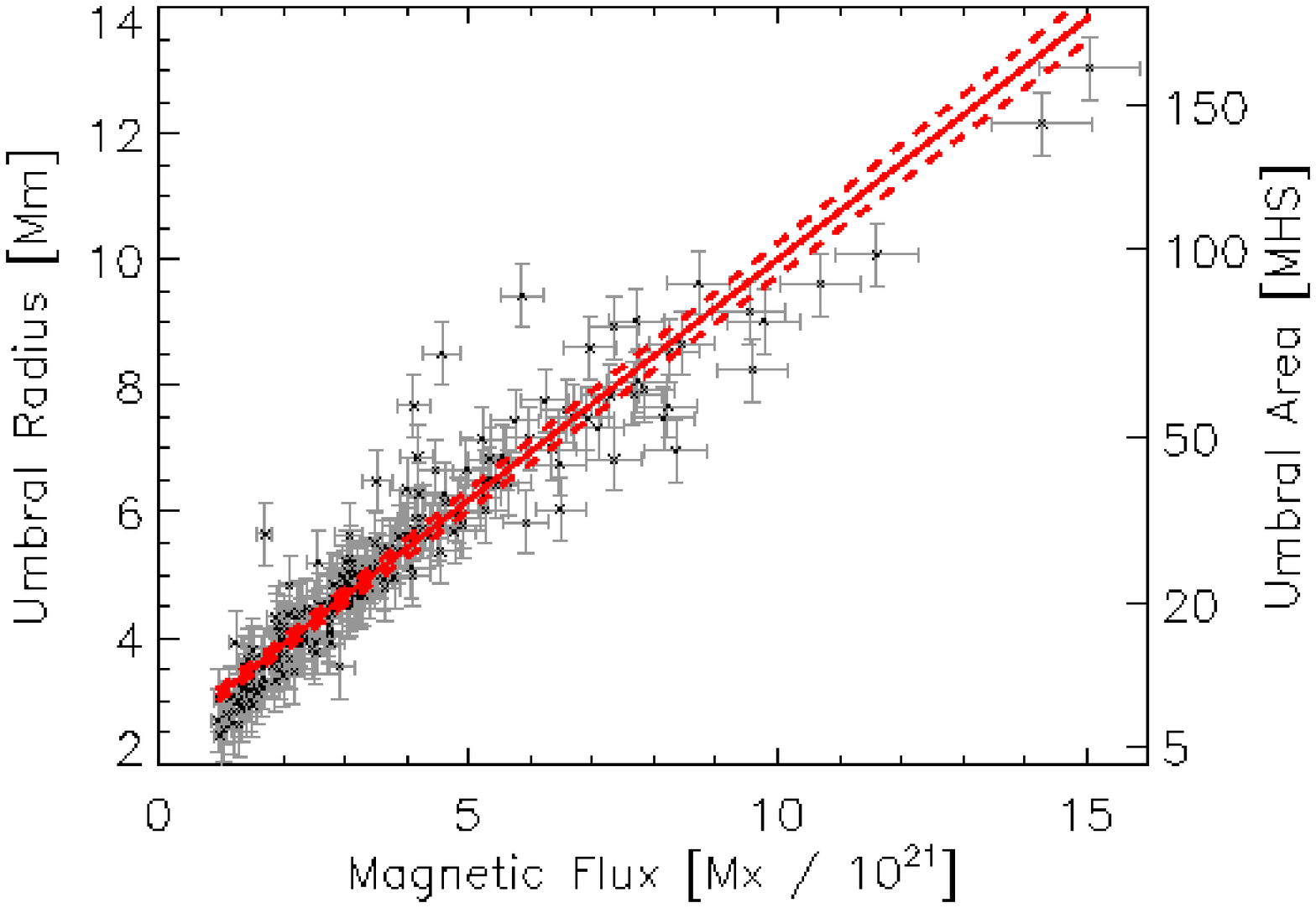}
  \includegraphics[width = \linewidth]{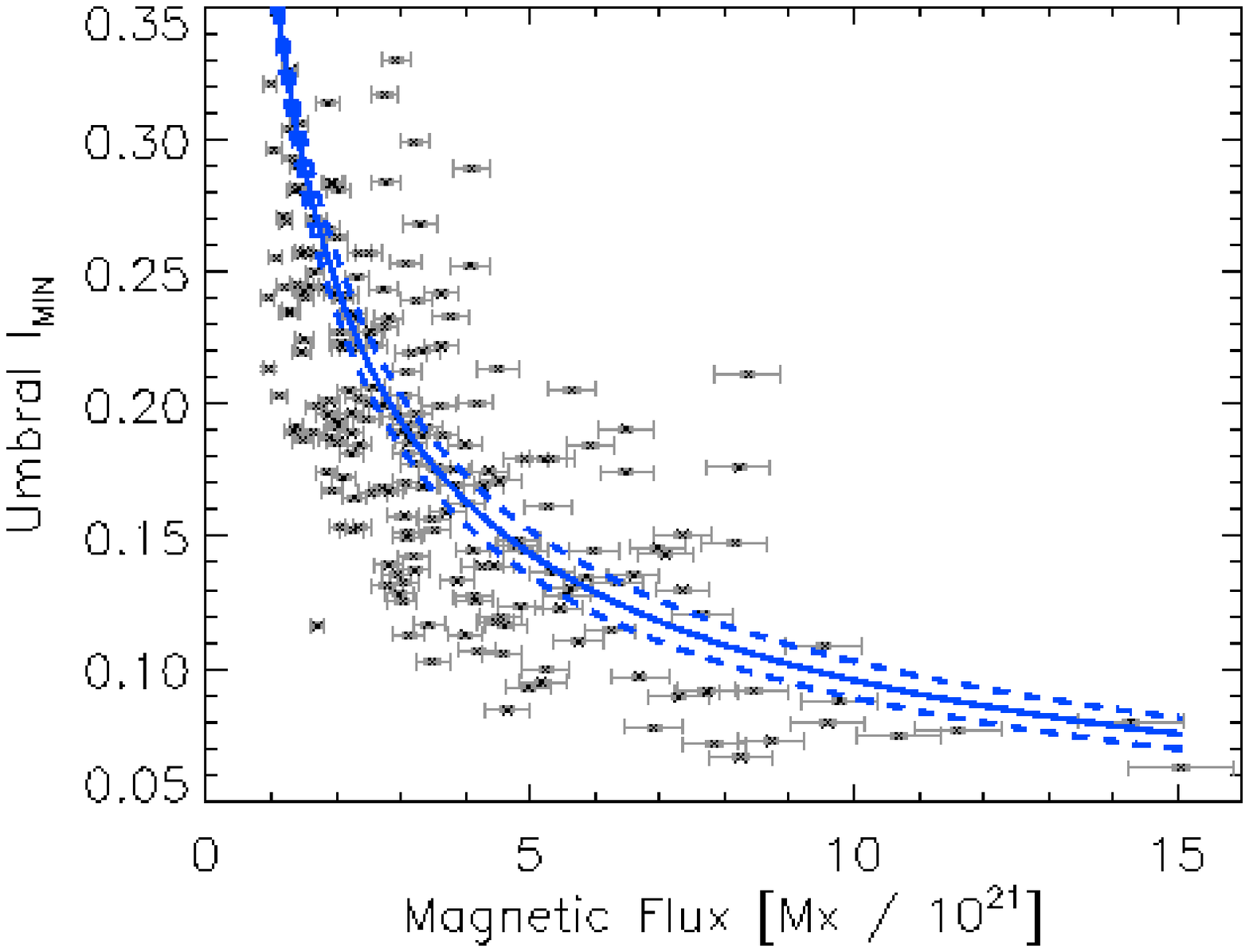}}
  \caption{
    Variation of umbral parameters with magnetic flux ($\Phi$). \emph{Left}: maximum field strength vs. $\Phi$, 
  \emph{center}: radius vs. $\Phi$, \emph{right}: minimum intensity vs. $\Phi$.
  Parameters of a linear fit (solid-red line) and power-law fit (solid-blue line) and correlation coefficients are 
  given in Table\,\ref{tab:result1}.
 }
  \label{fig:flux}
\end{figure*}
\subsection{Magnetic Flux}
We use the inclination maps from the inversions (Sect.\,\ref{sec:inversion}) 
to measure the total (unsigned) magnetic flux $\Phi$ of each sunspot in the small sample. 
Fig.\,\ref{fig:flux} shows scatter plots of the umbral size, 
intensity and field-strength versus flux. Maximum field strength and minimum intensity show nonlinear relations with the flux. 
The total magnetic flux of a sunspot is an integral quantity like the size.  
The umbral equivalent radius shows a quite linear dependence with flux (red line) and has a tight correlation (middle panel). 
The field strength and intensity show larger scatter compared to size and non-linear behavior. 
We use power-law relations to fit this data (blue lines, left and right panels).

\subsection{Northern and southern hemispheres}
\begin{figure}[!h]
\centering
\resizebox{8.0cm}{!}{\includegraphics[width = \linewidth]{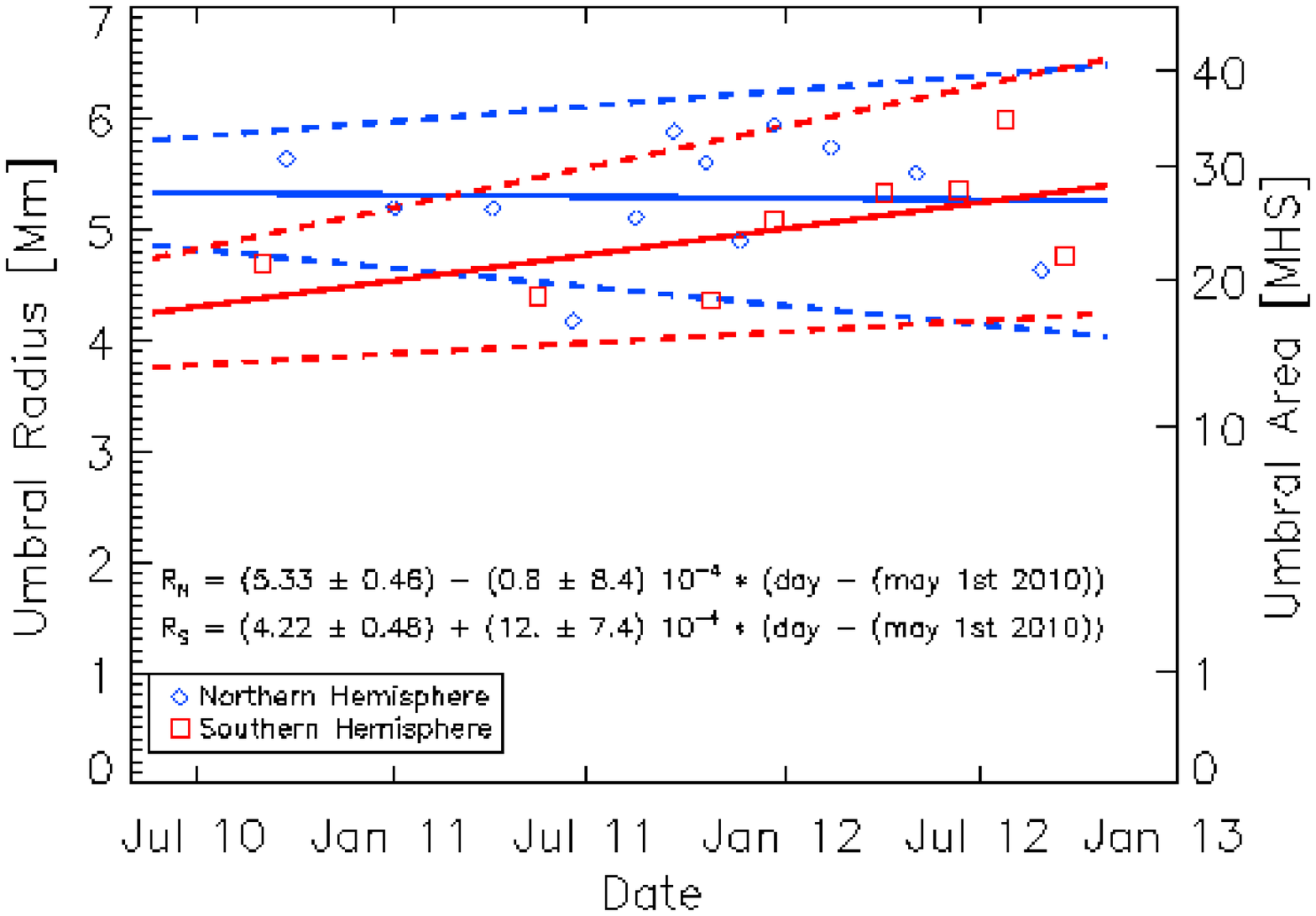}}
\resizebox{8.2cm}{!}{\includegraphics[width = \linewidth]{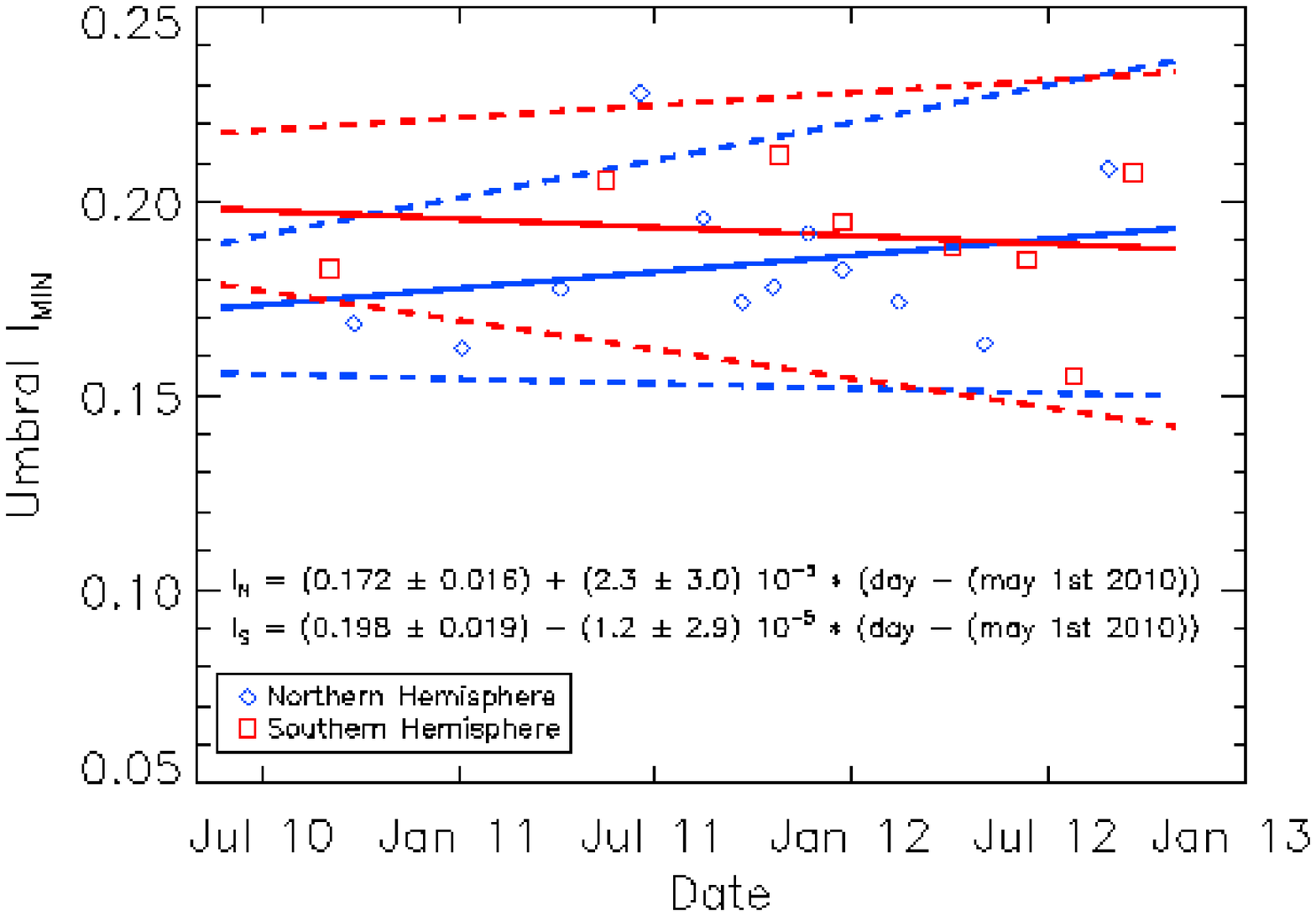}}
\caption{Variation of intensity and size of sunspots in each hemisphere vs. time. Each point represents the average of ten sunspots.}
\label{fig:hemisphere}
\end{figure}

There are about 50\% more sunspots in the northern hemisphere compared to the southern one in our full data (6892 umbrae). 
In a six-month interval statistics, the northern hemisphere had more sunspots than the southern except 
for the very last six month in our data. This might indicate a time lag between the two hemispheres. 
Therefore, we try to analyze the temporal variation of the radius and intensity of umbrae in each hemisphere (Fig.\,\ref{fig:hemisphere}). 
The northern hemisphere does not show a significant variation of the umbral radius vs. the solar cycle phase, i.e., between 2 years and 4.5 years 
after the beginning of the cycle, while there is a marginally significant change in the southern hemisphere ($2\sigma$). 
The temporal variation of the intensity (lower panel) does not show any significant variation in either hemisphere. 
This is an important finding and we come back to this point in Sect.\,\ref{sec:hemisphere_discuss}.

\subsection{Distribution of the umbral area}\vspace{-.2cm}
Both the sample of 910 unique umbrae (each counted once) and the whole dataset of 6892 umbrae were used to study the area distribution.
To approach the underlying probability density function (PDF), $\rho$,  
the histogram was normalized such that its area is unity. 
A Levenberg-Marquard least square procedure \citep[][chapter 15]{press_etal_92} was used to fit 
a lognormal function (Eq.\,\ref{eq:lognormal}) to this distribution. To weigh the bins, the 
errors were taken proportional to the inverse of the square root of 
the number of entries in each bin. 

\begin{figure}[!h]
\centering
  \resizebox{8.6cm}{!}{\includegraphics[width = \linewidth]{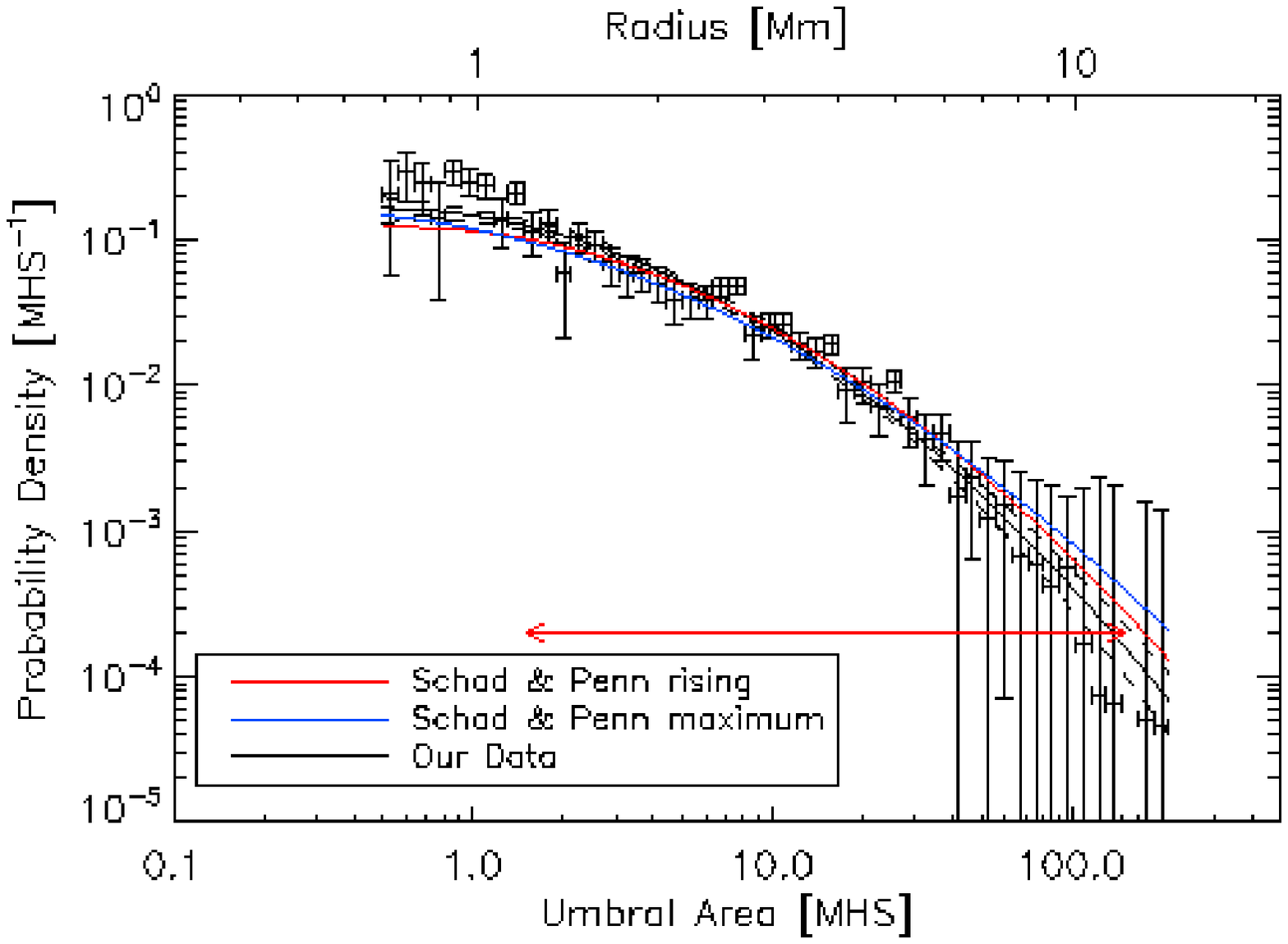}}
  \resizebox{8cm}{!}{  \includegraphics[width = \linewidth]{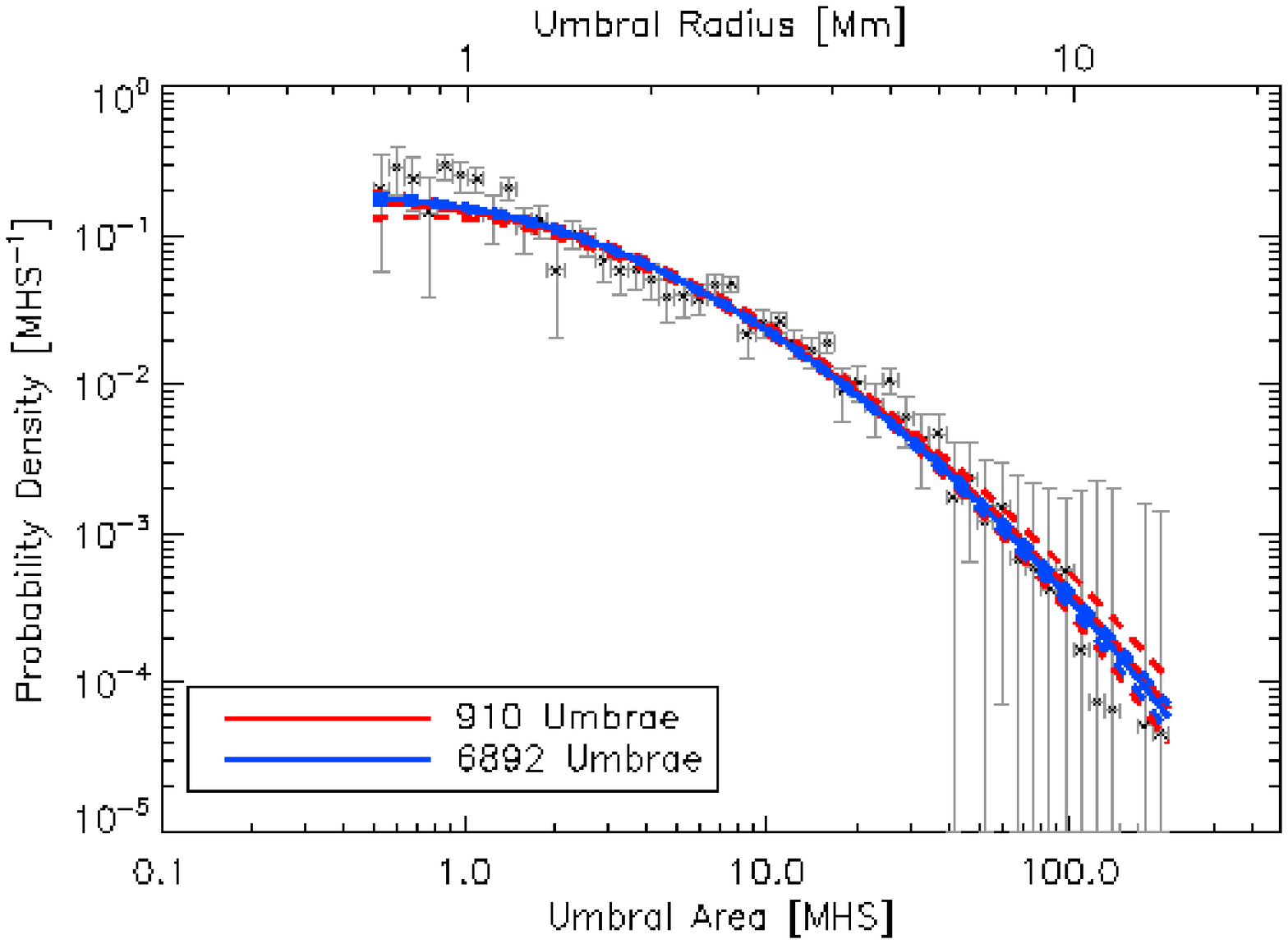}}
  \caption{Umbral size distribution. 
Error bars in $x$ direction mark the width of the bins. The error bars in $y$ direction are proportional 
to the inverse of the square root of entries in each bin. \emph{Top:} The black data points and curve show our data and the lognormal fit, respectively. 
The colored curves are adopted from SP and shows their fit (after normalizing its area) to the rising phase (red) and maximum (blue) 
of solar cycle 23. The red arrow marks the data range used by SP. 
\emph{Bottom:} Comparison of complete sample of 6892 umbrae and unique sample of 910 umbrae (no repetition). The dashed lines give the one sigma errors of the fit. 
}
  \label{fig:graph1}
\end{figure}

The lognormal distribution is 
\begin{equation}
\rho \left(x\right) = \frac{1}{\sqrt{2\pi}x\sigma}\exp\left(-\frac{\left(\ln x -\mu\right)^2}{2\sigma^2}\right),
\label{eq:lognormal}
\end{equation}
where $x$ is the umbral size in MHS, and $\mu$ and $\sigma$ are the two fit 
parameters. 
Table\,\ref{tab:results2} lists the fit parameters and their uncertainties. 
The mean value of the umbral area and the mean of the PDF in the observed range are 
the same (13\,MHS). The standard deviation of the data (20\,MHS) is similar to the value of the PDF (26\,MHS). 
 The top panel of Fig.\,\ref{fig:graph1} shows our data and the fit with uncertainties. 
The bottom panel of Fig.\,\ref{fig:graph1} shows the lognormal fits of the large sample and all data. 
The shapes of the two curves are very similar.
Neither counting each umbra only once, nor reducing the data to fully evolved sunspots drastically modifies the distribution compared to 
the full dataset.
It means that we do not see an influence of the evolution of sunspots on the distribution of umbral sizes in our data.

\begin{table}
\centering
\caption{Parameters of the lognormal fit and their one sigma errors. (Eq.\,\ref{eq:lognormal}).
The last two rows show fit parameters calculated from the parameters $\langle A \rangle$ 
and $\sigma_A$ (cf.\,Eq.\,\ref{eq:lognormal2}) published by 
SP and BOG.
}
\begin{tabular}[h]{lccccc}
\hline \hline
\multicolumn{1}{c}{phase} & \multicolumn{1}{c}{N} & \multicolumn{1}{c}{$\sigma$}  & \multicolumn{1}{c}{$\mu$}  \\\hline
large sample               & $910$   & $1.54\pm0.09$ & $1.64\pm0.12$ \\\hline
all data threshold 0.6     & $6892$  & $1.54\pm0.04$ & $1.56\pm0.05$ \\
all data threshold 0.5     & $5148$  & $1.56\pm0.05$ & $1.61\pm0.06$ \\
\hline
Schad \& Penn (rising phase) &  $5127$ & $1.60$        & $2.00$  \\
Bogdan etal (1917 - 1982)            & $24615$ & $1.16$        & $0.86$  \\
\hline
\end{tabular}
\label{tab:results2}
\end{table}

\section{Discussion}
\label{sec:discussion}
We performed a careful sunspot selection, taking mostly leading spots after the initial growth phase when most of the flux has emerged. 
About half of the selected sunspots have a $\mu$ value larger than 0.9 (heliocentric angle $\theta~<~26^\circ$). 
In the following, we discuss our results and compare them to earlier works.

\subsection{Relation between umbral size \& field-strength}

Already \cite{nicholson_1933} found that the magnetic field strength scales nonlinear with the umbral size,
although their results for field strengths below 2000\,G are biased by systematic errors.
Since then, this relationship was confirmed by several authors \citep[][]{kluber_1948, kopp_rabin_92, livingston_02, reza_etal_2012a}. 
In Fig.\,\ref{fig:bmax_size} we compare linear fits of SP (\ion{Fe}{i}\,868.8\,nm) and 
\citet[][\ion{Fe}{i}\,1.56\,\textmu m]{kopp_rabin_92} with our results. 
Since these spectral lines sample slightly different layers of the solar atmosphere compared to \ion{Fe}{i}\,617.3\,nm line (HMI), 
we evaluated the systematic offset in the received field strength. We used the COOL umbra 
model of \cite{collados_etal_94} and the SIR code \citep{sir92, sir_luis} to synthesize 
this model atmosphere for the given lines and invert them as in case of VFISV.
As expected, the \ion{Fe}{i}\,1.56\,\textmu m line samples deeper layers of the atmosphere and 
returns stronger field strengths (compared to HMI line) while \ion{Fe}{i}\,868.8\,nm line 
samples slightly higher layers, resulting in smaller values. Therefore, we applied a fixed correction (changing the ordinate) before 
over plotting the curves of SP (purple) and Kopp \& Rabin (yellow) in Fig.\,\ref{fig:bmax_size}. 
The line of SP is quite parallel to our line with an offset of about 300\,G.

\subsection{Size-intensity relation}
Figure\,\ref{fig:imin_size} shows a non-linear relationship between the minimum continuum intensity and the umbral radius. 
The umbral intensity contrast decreases with increasing wavelength. 
This complicates a direct comparison of the size-intensity and the intensity-field strength relation with 
other authors which is why we do not show adopted curves. This not only requires a correction factor to account for 
different wavelengths, but also requires individual correction for stray light removal applied in each study. 
A power-law fit (blue curve) 
matches the observed values better compared to a linear fit. 
SP presents a quadratic fit while \cite{mathew_etal_2007} find a power-law fit better reproduces the observations. 
The exponent of our power-law fit compares well with the exponent given by \cite{mathew_etal_2007}. These authors use a 
wavelength range close to the HMI line.

\subsection{Field strength-intensity relation}
It is generally believed that the 
sunspot umbra is darker due to a partial inhibition of convective energy transport in presence of 
a strong magnetic field \citep{biermann_41}. 
There is a well-known relation between the field-strength and intensity of umbrae \citep[e.g.][]{martinez_vazquez_1993}. 
Both, maximum field strength and minimum intensity show a non-linear dependence on size of umbrae with noticeable  
intrinsic scatter. 
The darkest umbrae in our data have minimum continuum intensities as low as six percent of quiet Sun intensity 
which indicates a low amount of stray light (Fig.\,\ref{fig:imin_b}). 
SP fit a power-law function to the field strength-intensity relationship while \cite{reza_etal_2012a} used a linear relation only. 
The power-law exponent in our fit (-0.26) is smaller than the one presented by SP because of the smaller umbral contrast 
in the near infrared.
The correlation between minimum intensity and maximum field strength is significantly higher than the one between size and intensity. 
Therefore, if no polarimetric measurements are available, the umbral intensity is a better proxy for a rough estimation of the 
magnetic field strength than the size as suggested by \cite{norton_gilman_2004}.

\subsection{Magnetic flux}
The general picture of a forming active region is the rise of a buoyant flux tube in the 
convection zone \citep{parker_1955,spruit_1981,cale_etal_1995}. The buoyancy requires 
definite values of the magnetic field strength, and 
rise of a buoyant flux tube is accompanied by a horizontal expansion. 
\cite{cheung_etal07}
find in simulations that a scaling relation forms between 
the field strength and the plasma density prior to the emergence on the solar surface. 
The final field strength 
is the result of the interaction between the rising flux tube and the turbulent convection in an stratified atmosphere. 
The total magnetic flux is one of the parameters determining the size, intensity and field-strength of a sunspot. 
During the formation process, pores and small flux patches migrate and accumulate enough flux to form sunspots 
\citep{zwaan_1992, leka_skumanich_1998, rolf_etal_2010b}. 
The total flux of a sunspot is less than the total flux of the corresponding flux tube since a fraction of flux disperses in 
plage and cancellations \citep[e.g.,][]{cheung_etal10}. 

For a better understanding of the interplay between the flux and maximum field strength as well as size and intensity, 
we plotted these parameters vs. flux in Fig\,\ref{fig:flux}. 
The scatter in the plot of field strength-flux (left panel) is reminiscent of the scatter in field strength-size 
plot (Fig.\,\ref{fig:bmax_size}). The umbral field strength perhaps also depends on how much flux goes into the penumbra. 
\cite{steinegger_etal_1990} find that the ratio of umbra to penumbra area has a positive correlation with the spot size. 
The linearity of the relation of sunspot radius vs. flux (Fig.\,\ref{fig:flux}, middle panel) suggests to use the sunspot 
size as a proxy for its flux. From this point of view the left and right panel of Fig.\,\ref{fig:flux} are similar to  
Figs.\,\ref{fig:bmax_size} and \ref{fig:imin_size}, respectively. 
Umbral sizes oscillate during the lifetime of spots and in particular for long-lived spots with a timescale of 3-5 days \citep{robinson_boice_82}. 
This is much longer than the dynamical timescale: the Alfv$\acute{\rm e}$n and sound timescale (at $\tau\!=\!1$ level) in a moderate-size spot 
is of the order of an hour \citep{rempel_schlichenmaier_2011}. 
It should be investigated if such size oscillations leave traces in the maximum field strength of spots.

\subsection{Maximum field strength and strength of a cycle}
At first glance, Fig.\,\ref{fig:bmax_size} suggests that the maximum field strength of sunspots level 
off at a certain value. 
In principle, the maximum field strength is a result of the horizontal pressure balance of umbrae and the 
surrounding atmosphere \citep[e.g.,][]{ossendrij_2003}. 
This simple picture, however, cannot explain the scatter in the left panel of Fig.\,\ref{fig:flux}. 
As seen in this figure, the maximum field strength does not increase much from sunspots with a flux of 5 
 to 15\,$\times$\,10$^{21}$\,Mx ($\approx$\,3.2\,kG and $\approx$\,3.4\,kG, respectively). 
\citet[][their Fig.\,2]{livingston_02} and \citet[][their Fig.\,3]{reza_etal_2012a} also do not find field strength 
larger than about 4\,kG in cycles 22 and 23, respectively. 

The strongest umbra in cycle 24 in our data has a field strength of 3.4\,kG,  
comparable to cycle 23 \citep[3.6\,kG using infrared lines,][]{reza_etal_2012a}. 
Our results are in agreement with \cite{norton_etal_2013} 
who find no variation in the maximum field strength of umbrae between cycle 23 and 24. 
\cite{livingston_etal_06} compiled old observations and report maximum field strengths up to 6\,kG. The number of sunspots with a very strong 
field strength, however, dropped sharply in the last few cycles. Does the Sun fail to generate large and strong sunspots? 
\cite{livingston_etal_06} find a fraction of 0.2\,\% for umbrae with a field strength larger than 4\,kG. 
In our sample of 205 spots we would expect 0.4 spots and do not observe such a strong field strength. 

The strength of solar cycle is determined by the number of sunspots, their darkness, and field strength. 
\cite{cameron_schuessler_2012}  propose that the strength of a cycle is changing as a function of 
cross-equator flux transport and near surface flows  \citep[see also ][]{durrant_turner_wilson_2004, cameron_etal_2013}. They use the 
open flux parameter \citep{wang_sheeley_2009} as a precursor of the next solar cycle. These authors find that 
the non-linear growth of disturbances due to an occasional emergence of a large active region near the solar equator 
can significantly amplify or weaken the next solar cycle. Hence it is perhaps too early to conclude that 
no sunspot appears in cycle 25 as proposed by \citet{livingston_etal_2012}.

\subsection{Hemispheric asymmetry and temporal variation}
\label{sec:hemisphere_discuss}
We present the temporal variation of intensity and size of umbrae in Fig.\,\ref{fig:hemisphere} and find no significant trend for 
the umbral intensity in either hemisphere during the rising phase of cycle 24. 
Our finding is in contrast to \cite{norton_gilman_2004} who find a decrease in umbral intensity in the rising phase 
of cycle 23. These authors used MDI instrument with a spatial resolution of about 2$\arcsec$. 
In comparison, HMI data has a spatial resolution of 1$\arcsec$. No stray light correction was applied in both studies.
\cite{mathew_etal_2007} find a decrease in 
the umbral intensity in the northern hemisphere but attribute it to variation of umbral size in that time interval and consider it insignificant. 
The time interval of our data is about half of the \cite{mathew_etal_2007}. 
Our finding is also in agreement with \cite{norton_etal_2013} who find no variation in umbral intensity during the rising phase of cycle 24, 
as well as \citet{Toma_2013} who find no variation in the umbral intensity from 1986 to 2012.

In our data the number of sunspots in the northern hemisphere is about 50\,\% larger than in the southern one. 
This either indicates that there is a time lag between the two hemispheres or there is a 
significant difference regardless of the phase of the solar cycle \citep{knaak_etal_2004}. There are several reports of 
the hemispheric asymmetry in meridoinal flow \citep{howard_gilman_1986}, 
polar field reversal \citep{durrant_wilson_2003}, and the magnetic activity \citep{temmer_etal_2002, brajsa_etal_2005}. 
\cite{mcintosh_etal_2013} find that overall in cycles, the northern hemisphere experienced a larger number of spots 
since 1965 while the southern hemisphere had an excess in the sunspot number in the declining stage of the last four cycles. 
Hence, our finding of a leading northern hemisphere is in agreement with the latter authors. 
An increase of the umbral radius is seen only in spots in the southern hemisphere (Fig.\,\ref{fig:hemisphere}).

\subsection{Umbral Size Distribution}

\subsubsection{Comparison with BOG and SP}\vspace{-.2cm}
To compare our result for cycle 24 with those of previous cycles, 
one has to convert the fit parameters in SP and BOG work to 
our definition. These authors use a different definition of the lognormal function,
\begin{equation}
  \ln\left(\frac{\textrm{d}N}{\textrm{d}A}\right) = 
  - \frac{\left(\ln A - \ln \langle A \rangle\right)^2}{2 \ln \sigma_A} +
  \ln\left(\frac{\textrm{d}N}{\textrm{d}A}\right)_\textrm{max},
  \label{eq:lognormal2}
\end{equation}
where $A$ is the umbral size, $({\textrm{d}N}/{\textrm{d}A})$ is the density function,
 $\langle A \rangle$ the mean and $\sigma_A$ the width of the distribution. 
With this definition, the function is normalized to its maximum rather than its 
integral ($\left(\frac{\textrm{d}N}{\textrm{d}A}\right)_\textrm{max}$ scales the distribution). 
To normalize Eq.\,\ref{eq:lognormal2} to its area, i.e., make it a PDF, one has to select 
$\left(\frac{\textrm{d}N}{\textrm{d}A}\right)_\textrm{max} = (2\pi \langle A \rangle^2 \sigma_A \ln \sigma_A )^{-1/2}$. 
It is straightforward to show that these two definitions are mathematically identical. 
The conversion factors are the following:
\begin{equation}
\sigma = \sqrt{\ln\sigma_A}\,\hspace{1cm}\rm{and}\,\hspace{1cm}\mu =\ln \langle A \rangle + \ln \sigma_A . 
\label{eq:conv}
\end{equation}
These relations allow to convert the fit parameters given by 
SP and BOG to our definition (Table\,\ref{tab:results2}). 

The top panel of figure\,\ref{fig:graph1} shows the lognormal fit of the umbral size distribution in the present 
solar cycle (black) and the previous one (red). The red curve was calculated 
using the results of SP for the ascending phase of solar cycle 23. 
We converted their parameters to our definition using Eq.\,\ref{eq:conv}. 
The curves of the rising part of solar cycle 23 
(red curve) and 24 (black curve) are similar 
(the fit parameters of these two fits are listed in Table\,\ref{tab:results2}.).
Although our fit parameters differ from the ones published by SP, the distributions are close 
in the relevant part of the size spectrum (i.e., between 0.5 and 200\,MHS where we actually fitted the data) and the largest deviation 
belongs to the smallest umbral radii.
This is also in agreement with an invariant umbral area distribution reported by BOG and SP. 
Therefore, we conclude that the umbral size distribution of the current solar cycle (24) 
is similar to the distribution of the previous one. 
Note that BOG did not use a threshold method to determine the umbral size, 
which makes the results not exactly comparable. 
Also note that they considered only spots within $\pm\,7.5$\,deg from the meridian while in our case as well 
as SP, all sunspots except those at the very limb are included. 

For umbrae larger than 100\,MHS there is a deviation between the fit and the data 
(Fig.\,\ref{fig:graph1}), which is also the case for cycle 
23 SP (their Fig.\,1). Although in both cases this difference is within 
the given errors, one might argue that the assumption of a lognormal distribution 
fails in this part of the size spectrum in the early stage of the cycle. 

\subsubsection{Influence of the intensity threshold}
\label{sec:threshold}
The continuum intensity at the umbra-penumbra border was assumed to be 0.6. For a discussion of the influence of this on 
the umbra-penumbra area ratio, see \cite{steinegger_etal_1996} and \cite{gyori_1998}.
We repeated our analysis for several intensity thresholds between 0.5 and 0.6 of quiet Sun mean intensity. 
Table\,\ref{tab:results2} contains the fit parameters for the umbral size distributions for thresholds of 0.5 and 0.6 I$_{\mathrm{c}}$.
The size distribution is very robust with respect to this parameter, the distributions
do not show a significant difference. 
We chose a lower threshold (0.52) for the analysis of umbral properties in the 205 sunspots for which we inverted the magnetic field.
This decision was made to reduce the risk of contamination of the umbra with penumbral light.

\subsubsection{Temporal invariance of the size distribution}
Our umbral area distribution (Fig.\,\ref{fig:graph1}) is similar to SP. 
These authors compare different parts of cycle 23 and find small variation in the umbral size distribution. 
BOG note that the distribution function does not change from one cycle to another. 
It is in contrast to other umbral properties like umbral brightness \citep{albregtsen78, albergtson_etal_1984} 
or magnetic field strength \citep{livingston_etal_2012,  reza_etal_2012a} which vary within the solar cycle, 
or show a long-term trend \citep{nagovit_rtal_2012}.

\subsection{Error discussion}
Intensity measurements always have to be taken with caution since they are to some degree biased by stray light. 
We do not correct for this since when we carried out the data analysis there was no point spread function (PSF) available for HMI.
Fortunately the low minimum umbral intensities we find in this work indicate a low degree of stray light contamination, 
or at least a rather narrow PSF compared to the size of large umbrae. 
See \citet{yeo_2013} for recent results regarding the PSF of HMI.

The errors to the curve fits in the scatter plots between intensity, magnetic field strength, radius and magnetic flux 
(Figs.\,\ref{fig:bmax_size}\,--\,\ref{fig:flux})
were calculated using the errors of the individual measurements, which are the spatial resolution of the telescope for the radius and a uncertainty of 150\,G for the magnetic field strength. Our findings can still be biased by random errors, especially since the statistics covers only 2.5 years of data. 
Therefore the relations for large spots have to be taken with caution, since the intrinsic scattering in all parameters is high.

The same argument is valid for the umbral size distribution as well. Our histogram falls below the lognormal curve for umbrae larger than 100\,MHS, 
but since the number of measurements in this part of the histogram is low, this may just be by chance since our statistics covers only 2.5 years. Due to the limited statistics the errors for the temporal variation of umbrae are large and an observed trend is 
barely significant.
On the other hand our statistics are, although short, complete and cover all sunspots during the rising phase of the current cycle (24).

\section{Summary and conclusion}
We use HMI data to investigate properties of sunspots in the 
ascending phase of the solar cycle 24. We use one solar image per day and 
apply an automated thresholding method to measure the umbral area. 
The data was corrected for limb-darkening and foreshortening effects.
We performed an inversion to retrieve the magnetic field vector for a hand-selected subset of 205 sunspots. 
We use the large sample (910 unique umbrae) to analyze the distribution of umbral sizes while the small sample 
enables us to study empirical relations and temporal variations of size, magnetic field strength, and continuum intensity.
The relations between magnetic field strength, continuum intensity and size of the umbrae 
are similar to previous studies, 
an invariant size-field strength relationship is likely to exist.

The relation between umbral intensity and size shows a power-law behavior 
with an exponent similar that of the previous cycle. 
Compared to a linear fit, a power-law function fits better the dependency of intensity on magnetic 
field strength and size. The continuum intensity is a good proxy of the field strength while sunspot flux 
closely follows the umbral area. 
The umbral radius shows an increase in the southern hemisphere, 
while no significant temporal variation in the northern hemisphere was observed. 
The umbral intensity does not show a temporal variation in the ascending phase of cycle 24 similar to cycle 23. 
The umbral size distribution is also similar to the one in the rising phase of cycle 23.
Using sunspot properties in ascending phase of cycle 24, 
we do not find any evidence for a significant decrease in solar activity as suggested by e.g., \citet{livingston_etal_2012}. 
We will continue to monitor sunspot properties and their changes through the rest of this cycle, 
to examine the likelihood of a long-term decrease of the solar activity.

\begin{acknowledgements}
The data used here are courtesy of NASA/SDO and the AIA, EVE, and HMI science teams. 
R.R. acknowledges financial support by the DFG grant RE 3282/1-1.  
We acknowledges fruitful discussions at the workshops on ``Filamentary Structure and Dynamics of 
Solar Magnetic Fields'' as well as  ``Extracting Information from spectropolarimetric observations: comparison of inversion codes'' 
at the ISSI in Bern. We thank Rolf Schlichenmaier for insightful discussions. 
We also thank the anonymous referee for helpful suggestions. 
We are grateful to Juan Manuel Borrero for instructions on using his latest version of VFISV code. 
We use data provided by M. Rempel at the National Center for Atmospheric
Research (NCAR). The National Center for Atmospheric Research is sponsored 
by the National Science Foundation. 

\end{acknowledgements} 

\bibliography{christoph}

\end{document}